\documentclass[12pt]{iopart}
\usepackage{iopams}

\usepackage{graphicx}
\usepackage{dcolumn}
\usepackage{bm}
\usepackage{dsfont}
\usepackage{amssymb}
\usepackage{sidecap}
\usepackage{wrapfig}

\usepackage{graphicx}
\usepackage{leftidx}
\usepackage{color}

\newcommand{\ket}[1]{\left| #1 \right>} 
\newcommand{\bra}[1]{\left< #1 \right|} 
\newcommand{\vev}[1]{\langle #1 \rangle}

\newcommand{\correction}[1]{{\color{black} #1}}

\def\beq{\begin{equation}}
\def\eeq{\end{equation}}
\def\beqa{\begin{eqnarray}}
\def\eeqa{\end{eqnarray}}

\begin{document}

\title[Crystalline structures and frustration in a two-component Rydberg gas]{Crystalline structures and frustration in a two-component Rydberg gas}

\author{Emanuele Levi, Ji\v{r}\'{i} Min\'{a}\v{r}, Juan P. Garrahan and Igor Lesanovsky}
\address{School of Physics and Astronomy, University of Nottingham, Nottingham, NG7 2RD, United Kingdom}

\begin{abstract}
We study the static behavior of a gas of atoms held in a one-dimensional lattice where two distinct electronically high-lying Rydberg states are simultaneously excited by laser light. We focus on a situation where interactions of van-der-Waals type take place only among atoms that are in the same Rydberg state. We analytically investigate at first the so-called classical limit of vanishing laser driving strength. We show that the system exhibits a surprisingly complex ground state structure with a sequence of compatible to incompatible transitions. The incompatibility between the species leads to mutual frustration, a feature which pertains also in the quantum regime. We perform an analytical and numerical investigation of these features and present an approximative description of the system in terms of a Rokhsar-Kivelson Hamiltonian which permits the analytical understanding of the frustration effects even beyond the classical limit.
\end{abstract}

\maketitle


\section{Introduction}
Atoms in highly excited states --- so-called Rydberg atoms --- interact via power-law potentials, which in conjunction with an external laser drive give rise to intricate many-body phenomena. Recent experiments with Rydberg gases have revealed that the dynamical behavior of these systems is of surprising variety. Examples are the emergence of bistable behavior \cite{Carr2013,Malossi2014}, the observation of correlated aggregation of excitations \cite{Schempp2014} and of coherent excitation transport \cite{Ditzhuijzen08,gunter_observing_2013,Barredo14}. Strong interactions also become manifest in the structure of the ground state of Rydberg ensembles. Several theoretical works have predicted and studied the formation and the melting of crystalline phases \cite{Schachenmayer10,Pohl10,vanBijnen11,Sela11,Lesanovsky12,Ji,Honing13,Weimer,Vermersch15}. But only recently it was shown that these crystalline states can be actually accessed experimentally \cite{Schauss14}.

Currently there is a surge of interest in atomic systems in which multiple Rydberg states are excited. One of the main motivations is the presence of an exchange interaction between Rydberg states of different atoms that results in coherent transport dynamics \cite{Ditzhuijzen08,gunter_observing_2013,Barredo14}, the non-local propagation of light \cite{Li14} as well as a non-trivial collapse and revival dynamics \cite{Bettelli13,Maxwell14}.
Interesting physics emerges also in the absence of exchange interactions. The main reason is that depending on the Rydberg states the interaction between two atoms can vary by one or more orders of magnitude \cite{Saffman_2010,Beguin_2013}. This feature is exploited for example in all-optical transistors for light pulses \cite{Gorniaczyk14,Tiarks14}.

In this work we explore the many-body ground states of an atomic lattice gas in which two Rydberg states --- or species --- are simultaneously excited. We focus on a situation where interactions are present only between atoms of the same species. This is reminiscent of the Potts model \cite{Wu_1982}, for which however only few studies exist that consider interactions that extend beyond nearest neighbors \cite{Herrmann_1984, Glumac_1993, Cannas_1997, Bayong_1999}. We show that the ground state of the two-species Rydberg lattice gas features a surprisingly complex structure with a series of compatible to incompatible transitions. In the compatible case the two species can occupy the lattice such that they can both minimize their interaction energy independently. In the incompatible case this is not possible and it leads to mutual frustration. We provide analytical expressions for the regions of stability of the (in)compatible phases in the classical regime, i.e. in the limit of vanishing laser driving strength. The quantum regime is studied numerically as well as analytically with the help of an approximate Hamiltonian of Rokhsar-Kivelson form. These considerations show how frustration persists also in the quantum regime.

\begin{figure}[t]
\centering
	\includegraphics[trim = 0mm 0mm 0mm 0mm, clip, width=\textwidth]{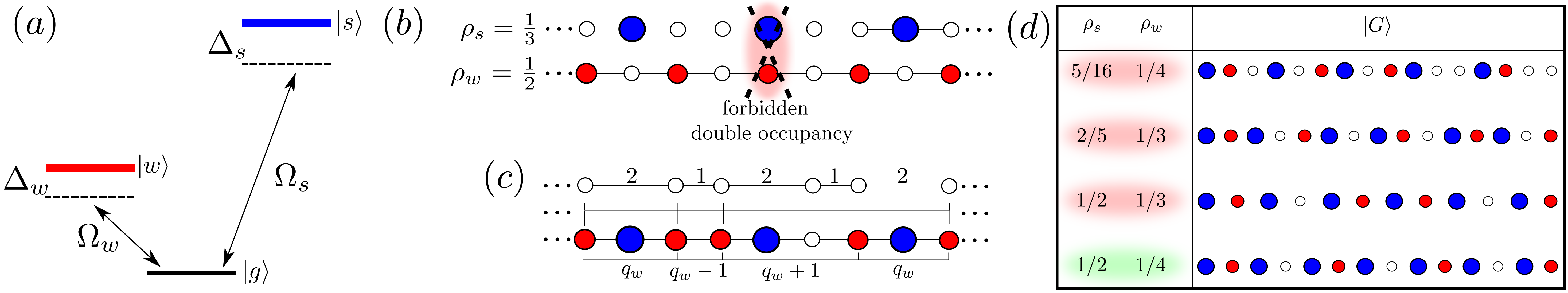}
\caption{(Color online) (a) Relevant atomic levels. Two Rydberg nS-states $\ket{\alpha}$, $\alpha=s,w$ (referred to as species) are excited from the ground state $\ket{g}$ by a laser with Rabi frequencies $\Omega_{\alpha}$ and detuning $\Delta_{\alpha}$. Atoms in the state $\ket{s}$ ($\ket{w}$) interact strongly (weakly). (b) Sketch of two minimum interaction-energy configurations for the strong and weak species at incompatible filling fractions. (c) Achieving a minimum energy disposition. The available lattice sites after placing the $s$-species form a distorted lattice with spacings of 1 and 2 times the original lattice spacing. Atoms of the $w$-species are placed in the available lattice sites according to the algorithm of Ref. \cite{Hubbard}. (d) Examples of ground states with different filling fractions $\rho_\alpha$. Incompatible (compatible) cases are shaded in red (green).}
\label{fig:1}
\end{figure}

\section{Description of the system} 
\begin{figure}[t]
\centering
	\includegraphics[trim = 0mm 0mm 0mm 0mm, clip, width=0.6\textwidth]{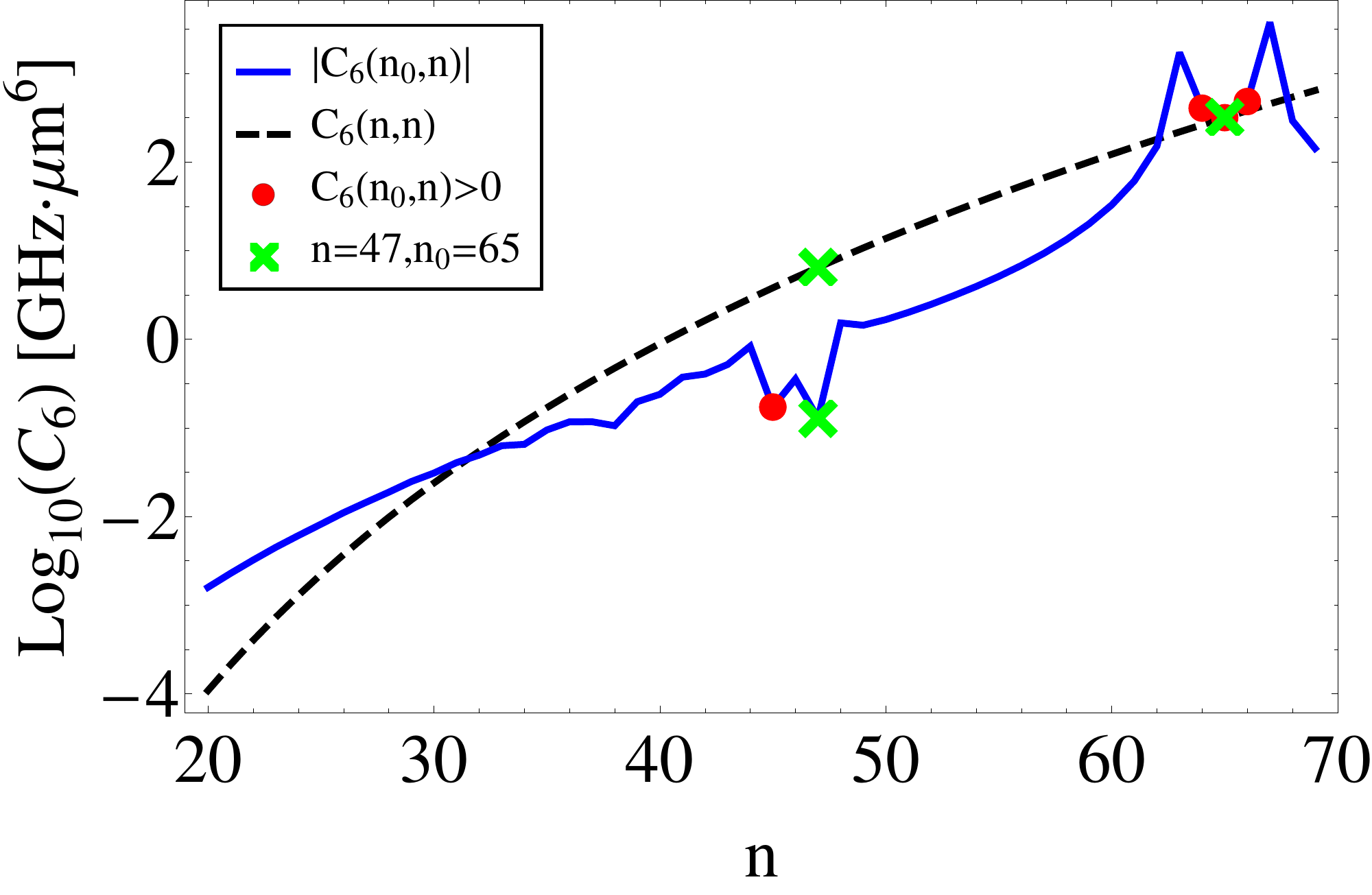}
\caption{(Color online) Comparison of the van-der-Waals $C_6(n_0,n)$ coefficients describing the interaction strength between two rubidium atoms in Rydberg states with principal quantum number $n_0$ and $n$ respectively. $C_6$ for the inter-species case is shown as a blue line (attractive interaction) and red circles (repulsive interaction). The black dashed line corresponds to the intra-species $C_6$, i.e. $n_0=n$. Green crosses represent a possible choice of Rydberg levels implementing the required hierarchy of interaction strengths assumed in our work. Identifying $n_0=65$ with the strong species and $n=47$ with the weak one we get $C^{(s)}_6/C^{(w)}_6 = 50.5$ and $C^{(w)}_6/C^{(sw)}_6 = 49$, i.e. the hierarchy $C_6^{(s)} \gg C_6^{(w)} \gg C^{(sw)}_6$ is well respected (we have denoted here by $C_6^{(sw)} = C_6(65,47)$ the inter-species $C_6$ coefficient).
}
\setlength{\unitlength}{1cm}	
  	\begin{picture}(2,2)
		  \put(4.4,11.7){\normalsize{ $C_6^{(s)}$ }}
		  \put(1.5,11.7){\normalsize{ $C_6^{(w)}$ }}
		  \put(1.5,9.7){\normalsize{ $C_6^{(sw)}$ }}
	\end{picture}
\label{fig:X}
\end{figure}
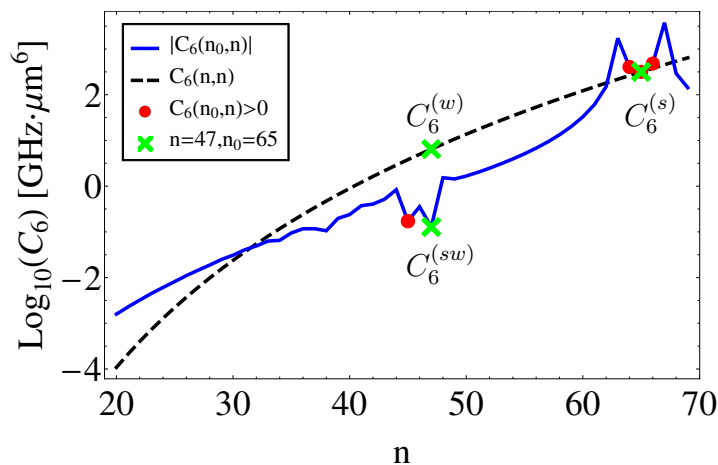

We consider a one-dimensional gas of atoms trapped in a lattice of spacing $a$ with a single atom per site. The relevant internal level structure of the atoms is shown in Fig. \ref{fig:1}a. The ground state $\left|g\right>$ is coupled to two Rydberg nS-states ($\left|s\right>$ and $\left|w\right>$) via two laser fields with Rabi frequencies $\Omega_\alpha$ and detuning $\Delta_\alpha$ ($\alpha=s,w$). The fundamental interaction between two such atoms is to leading order given by the dipole-dipole potential. However, for Rydberg atoms in nS-states, as considered here, this interaction results in a (second order) van-der-Waals energy shift \cite{Saffman_2010}. For two atoms in the pair state $\left|\alpha\alpha\right>\equiv\left|\alpha\right>\left|\alpha\right>$ and separated by the distance $R$ it reads $V_{\alpha}(R) = C^{(\alpha)}_6\,R^{-6}$. The van-der-Waals coefficients $C^{(\alpha)}_6$ scale with the eleventh power of the principal quantum number $n$ \cite{Gallagher_1994,Saffman_2010,Rydberg2}. This generates strong interactions among Rydberg states and moreover permits to achieve a scenario in which the interaction between two atoms in the pair-state $\ket{ss}$ is much larger than the interaction between two atoms in $\ket{ww}$. Henceforth we focus on such a case, which is in practice achieved by choosing the principal quantum number of state $\ket{s}$ (the strong species) to be larger than that of state $\ket{w}$ (the weak species). Moreover, such large difference in the principle quantum numbers results in a strongly suppressed interspecies interactions \cite{olmos_2011} as experimentally shown in Ref. \cite{Teixeira}.
This is a rather generic feature of Rydberg atoms in the sense that one can find various combinations of suitable levels which satify such strong-weak hierarchy. Identification of these levels is atomic species dependent and the corresponding van-der-Waals coefficients can be calculated using perturbative techniques as described e.g. in \cite{Singer_2005}. To give an idea of the orders of magnitude of the $C_6$ coefficients and how they depend on the principle quantum number, we provide a typical example in Fig. \ref{fig:X}, where we compare the inter- and intra-species interactions for Rydberg states of rubidium.

The Hamiltonian of the system can be written (within the rotating-wave approximation) as the sum of Hamiltonians of the strong and weak species, $H=H_w + H_s$, with
\begin{eqnarray}
  H_{\alpha} &=& \sum_{k=1}^{L}\left[\Omega_{\alpha}\sigma^{x(\alpha)}_k-\Delta_{\alpha}n^{(\alpha)}_k+V_\alpha\sum_{l>k}\frac{n^{(\alpha)}_k n^{(\alpha)}_l}{\left(l-k\right)^6}\right], \label{eq:hamiltonian}
\end{eqnarray}
where $n^{(\alpha)}=\ket{\alpha}\bra{\alpha}$ and $\sigma^{x(\alpha)}=\ket{\alpha}\bra{g} + \ket{g}\bra{\alpha}$, and $V_\alpha=C_6^{(\alpha)}/a^6$.
Note, that $H_w$ and $H_s$ do not commute due to the fact that both Rydberg species are excited from a common ground state. This is the origin of the compatible to incompatible transitions as well as the frustration effects which we will discuss in the following.

\section{Compatible to incompatible transition}
We start by studying the case $\Omega_{\alpha}=\Delta_{\alpha}=0$ in the thermodynamic limit, i.e. an infinite lattice, where Eq. (\ref{eq:hamiltonian}) reduces to the Hamiltonian of a classical two-species Potts-model with (convex) $1/r^6$ interactions and inequivalent couplings ($V_s \neq V_w$). We are interested in finding the microscopic state which minimizes the interaction energy for given filling fractions of excitations $\rho_s$ and $\rho_w$. For a single species and a general convex potential this problem was studied in \cite{Pokrovsky,Hubbard,Bak,Burnell}. Given the filling fraction the convexity of the potential forces the system into the most homogeneous configuration achievable accounting for the lattice constraint, as shown by Hubbard in Ref. \cite{Hubbard}.

When two species are present it is in general not possible to minimize at the same time the interaction energy of both. An example is given in Fig. \ref{fig:1}(b) for $\rho_s=1/3$ and $\rho_w=1/2$. Finding the actual ground state is simplified by our assumption $V_{s} \gg V_{w}$. Here, the configuration of the $s$-species can be considered as ``frozen'' and not constrained by that of the $w$-species \footnote{This is strictly speaking true only in the asymptotic limit $V_s/V_w \rightarrow \infty$. Otherwise an additional assumption has to be made on the filling fractions, namely $V_s/V_w \gg 1$ \emph{and} $\rho_s/\rho_w > 1$, the situation we consider here. See also \cite{InPrep} for details.}. Atoms of the $s$-species will then arrange following the single species prescription of Ref. \cite{Hubbard}, that is calling $r_l$ the distance between an $s$-atom and its $l$-th nearest neighbor, the $s$-atoms will be distributed on the chain with distances satisfying $r_l=\left\lfloor l/\rho_s\right\rfloor$ or $\left\lceil l/\rho_s\right\rceil$.
This set of constrains identifies a unique distribution for the atoms of the $s$-species, and will in general lead to a deformed lattice formed by the remaining empty sites on which the minimum-energy arrangement of the atoms of the $w$-species needs to be found.

For the example shown in Fig. \ref{fig:1}(b) the $w$-atom sitting in the doubly occupied site has to be moved to an empty site in a way that minimizes the increment in interaction energy. As shown in Fig. \ref{fig:1}(c) the strong species leaves a distorted lattice with a density of empty sites given by $1-\rho_s=2/3$ and lattice spacings $a$ and $2a$. The minimum interaction energy is then obtained by arranging the $w$-atoms according to the single species prescription of Ref. \cite{Hubbard} considering that the filling fraction of the $w$ species on the distorted lattice is $\rho_w/\left(1-\rho_s\right)=3/4$. This finally leads to the state depicted in Fig. \ref{fig:1}(c). The mathematical proof of this method we provide in a separate publication, see Ref. \cite{InPrep}. Some examples of minimum interaction--energy configurations are shown in Fig. \ref{fig:1}(d). From these considerations one finds that two cases, which depend on the filling fractions $\rho_\alpha$, need to be distinguished: (i) The compatible case in which the two species assume their minimum--energy dispositions independently without interfering with each other. (ii) The incompatible case in which the strong species prevents the atoms of $w$-species from assuming their minimum energy configuration.

Understanding whether two filling fractions are compatible is generally a hard task which requires the study of the full arrangement of excitations. In the following analytical analysis we will consider filling fractions of the form $\rho_\alpha=1/q_\alpha$, where both $q_{\alpha}$ are positive integers. Clearly, the minimum energy configuration is achieved when the $\alpha$-atoms are arranged uniformly with distance $q_\alpha$. The filling fractions are compatible only if they share a common divisor (see \cite{InPrep}). In the incompatible case the distribution of the $w$-atoms will be distorted and can be represented by repeated strings of repeated intervals of lengths $...q_w(q_w-1)(q_w+1)q_w...$, as shown for the example in Fig. \ref{fig:1}(c).

Let us now turn to the discussion of the stability of the (in)compatible phases. So far we have assumed that the filling fractions $\rho_\alpha$ of the individual species are externally imposed. However, in practice the filling fraction is controlled by the parameters $\Delta_\alpha$ in Hamiltonian (\ref{eq:hamiltonian}) which act as chemical potentials \cite{Pohl10,Weimer}. The question is then which state, or filling fraction, is actually stabilized in a given region of the $\Delta_{s}-\Delta_{w}$ manifold. As the distribution of the $s$-species is effectively independent from that of the $w$-atoms, we can study its stability with the single species method \cite{Bak,Burnell,Weimer}. The stability of a given filling fraction $\rho_w$ on the other hand is less simple to analyze, as it depends on the configuration of the $s$-atoms. We can obtain an analytical result for the transition to an incompatible state in the thermodynamic limit starting from a compatible arrangement of atoms for which $\nu=q_w/q_s$ is integer (see Section 6 in \cite{InPrep}). The region of stability is delimited by
\begin{equation}
\label{eq:stability}
\frac{\Delta_w^{(\pm)}}{V_w}=\sum_{l=1}^{\infty}\left[\frac{1\mp l(q_w-\nu)}{(lq_w)^6}\pm\frac{l(q_w-2\nu)}{(lq_w\mp1)^6}\pm\frac{l\nu}{(lq_w\mp 2)^6} \right].
\end{equation}
On the one hand a transition to an incompatible state takes place when $\Delta_w=\Delta_w^{(+)}$, the value of $\Delta_w$ for which the introduction of one more excitation is energetically favorable when keeping $\Delta_s$ fixed.
In this case the distribution of the $l$-th nearest neighbor $w$-atoms which in the compatible phase is homogeneous, e.g. atoms at distance $lq_w$, is deformed by the introduction of $l\nu$ distances $q_w-2$, and $l(q_w-2\nu)$ distances $q_w-1$.
On the other hand there is a transition at $\Delta_w=\Delta_w^{(-)}$ at which the $w$ species loses an excitation and the compatible distribution is deformed by the introduction of $l\nu$ distances $q_w+2$, and $l(q_w-2\nu)$ distances $q_w+1$.


\begin{figure}[t]
\centering
	\includegraphics[trim = 0mm 0mm 0mm 0mm, clip, width=0.7\textwidth]{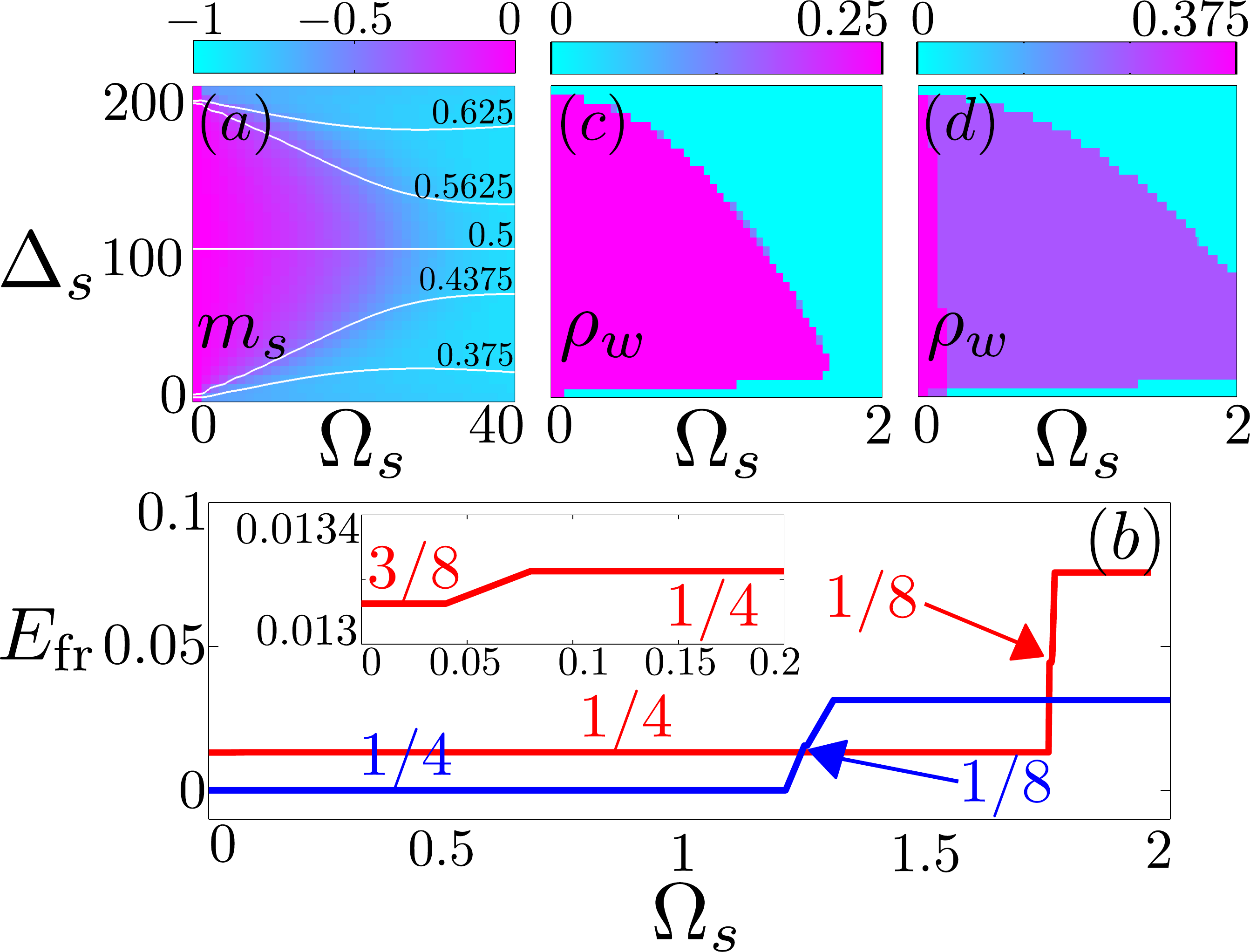}
\caption{(Color online) Exact diagonalization results. (a) Density plot showing the transverse magnetization of the strong species $m_s$ in a parameter regime $\Delta_s$ where the filling fraction in the classical limit $\rho_s^{\rm cl}$, i.e. at $\Omega_s=0$, is $1/2$.  The white contours show lines of equal $\rho_s$. (b) Frustration energy (\ref{eq:fe}) plotted along $\Delta_s=102$, i.e. the $\rho_s=1/2$ contour of (a). The blue and red lines correspond respectively to a compatible ($\rho^{\rm cl}_w=1/4$) and an incompatible ($\rho^{\rm cl}_w=3/8$) classical filling fraction. The value of $\rho_w$ is indicated explicitly in the figure for some plateaus. The inset magnifies the first small jump in $\rho_w$ in the incompatible cases. (c,d) Filling fractions of the weak species for compatible [$\rho^{\rm cl}_w=1/4$, (c)] and incompatible [$\rho^{\rm cl}_w=3/8$, (d)] case. The data were obtained with $V_s=100V_w=100.$ }
\label{fig:2}
\end{figure}

\section{Quantum fluctuations and frustration energy.}
We now address the question as to how quantum fluctuations introduced by a laser of finite driving strength affect the compatible to incompatible transitions and in particular the emergence of frustrated states. To this end we consider a non-zero value of $\Omega_s$. This changes the state of the $s$-species from a classical one to a superposition of configurations with different excitation number. We are interested in how this impacts the classical arrangements of the $w$-species, i.e. at $\Omega_{w}=0$. To quantify this we introduce the frustration energy
\begin{eqnarray}
 \label{eq:fe}
   E_{\rm fr} = E^{(w)}_{\rm 2 sp}(\Delta_{s,w};\Omega_{s}) - E^{(w)}_{\rm 1 sp}(\Delta_{w}),
\end{eqnarray}
which measures by how much the strong species prevents the $w$-species from reaching its minimum energy configuration if it were alone. Here $E^{(w)}_{\rm 2 sp}(\Delta_{s,w};\Omega_{s}) = \vev{ H_{w} }$, i.e. the expectation value of $H_w$ [see Eq. (\ref{eq:hamiltonian})] and $E^{(w)}_{\rm 1 sp}(\Delta_{w})$ is the energy of the classical configuration of the $w$-species in the absence of $s$-atoms.

In the classical limit, $\Omega_\alpha=0$, the frustration energy is zero when the detunings $\Delta_\alpha$ are chosen such that they stabilize filling fractions $\rho_\alpha$ which are compatible. $E_{\rm fr}$ becomes in general larger with increasing $\Omega_s$ as increasing density fluctuations in the $s$-species force the $w$-atoms to assume configurations with increased energy. In order to study this behaviour in more detail we diagonalize Hamiltonian (\ref{eq:hamiltonian}) for a chain of length $L=8$ with periodic boundary conditions. This approach has the typical drawbacks of a small scale numerical study: the presence of finite size effects (though minimized by the periodic boundaries), and the fact that one is limited to filling fractions of the form $p/L$ for $p\leq L$. We will see, however, that the results provide an intuitive understanding of the physics at work. Note furthermore, that the considered relatively small system size in fact comes close to what is currently realizable experimentally in the context of Rydberg atoms \cite{Schauss12,Schauss14}.

For our numerical study we focus on a regime where the strong species in the classical limit forms a crystal with filling fraction $\rho^\mathrm{cl}_s=1/2$. With increasing $\Omega_s$ this crystal melts as is seen in Fig. \ref{fig:2}(a) where we show a density plot of the transverse magnetization $m_s = (1/L) \sum_{k=1}^L\vev{\sigma_k^{x(s)}}$. The magnetization displays the typical lobe-structure \cite{Weimer}, and the formation of a (longitudinally) paramagnetic state ($m_s =-1$) at large $\Omega_s$. Here it is evident that the state of the strong species is formed by a superposition of states with different number of excitations.

Let us now have a look at the frustration energy when the weak species is added. The corresponding data is shown in Fig. \ref{fig:2}(b).
In the classical limit ($\Omega_s=0$) we choose $\Delta_s = 102$ which corresponds to half filling of the strong species, $\rho^{\rm cl}_s=1/2$, and consider values of $\Delta_w$ which stabilize the compatible case $\rho^{\rm cl}_w=1/4$ ($\Delta_w\simeq 0.0159$ blue line) and the incompatible case $\rho^{\rm cl}_w=3/8$ ($\Delta_w\simeq 0.0314$, red line), respectively. We first determine the values of $\Delta_w$ numerically as the midpoints of the stability regions. In the compatible case one can use Eq. (\ref{eq:stability}) to determine $\Delta_w$. In the incompatible case on the other hand Eq. (\ref{eq:stability}) cannot be applied directly.
In the finite size case though, as previously mentioned, only fillings of the form $p/L$ are accessible, and as such the region of stability of $\rho^{\rm cl}_w=3/8$ extends from the upper boundary $\Delta^{(+)}_w$ for $\rho_w=1/4$ to the lower boundary $\Delta^{(-)}_w$ for $\rho^{\rm cl}_w=1/2$. These cases are compatible such that we can use again Eq. (\ref{eq:stability}) to estimate the midpoint of this region of stability. We find that the values obtained analytically from Eq. (\ref{eq:stability}) are in remarkable agreement with the numerically determined values despite the small size of the system.
At $\Omega_{s}=0$, $E_{\rm fr}$ is zero for compatible densities and finite for incompatible ones. With increasing $\Omega_{s}$, the $w$-species becomes more and more frustrated which is reflected in an increase of $E_{\rm fr}$. Interestingly, this increase is step-wise and each step is accompanied by a change of the filling fraction $\rho_w$ of $w$-atoms, whose value is provided in Fig. \ref{fig:2}(b) for each plateau of $E_{\rm fr}$.

Furthermore, we show the behavior of the weak filling fraction in the entire $\Omega_{s}-\Delta_{s}$ plane in Fig. \ref{fig:2}(c,d). Note, that contrary to $\rho_s$ [shown as contours in Fig. \ref{fig:2}(a)], $\rho_w$ exhibits a lobe like structure which is not symmetric. This can be understood as follows: For a given finite $\Omega_{s}$ the filling fraction $\rho_s$ decreases with decreasing $\Delta_s$. The resulting smaller number of $s$-atoms permits the accommodation of a larger number of $w$-atoms. This effect is stronger the larger $\Omega_{s}$ leading to an increasing asymmetry of the lobes in Figs. \ref{fig:2}(c,d). Moreover, for any $\Delta_s$, $\rho_w$ decreases stepwise with increasing $\Omega_s$ and eventually vanishes. The decrease in $\rho_w$ can be qualitatively explained as follows: As $\Omega_s$ increases the ground state begins to contain basis states with $s$-atom numbers that are larger than in the classical case. This forces the $w$-atoms to assume a lower filling fraction. The fraction of admixed basis states with increased $s$-atom number becomes larger the larger $\Delta_s$ [see contours in Fig. \ref{fig:2}(a)] which results in the asymmetry of the lobes in Figs. \ref{fig:2}(c,d).

We expect the qualitative features, namely the increase of frustration with $\Omega_{s}$ to hold also in the thermodynamic limit. To study this a semi-analytical treatment of quantum fluctuations based on a perturbative expansion of the ground state around the classical limit could in principle be performed. This task is, however, quite involved due to the fact that the filling fraction of the weak species is in fact a function of $\Omega_{s}$. Nevertheless, one can still gain analytical insight into the quantum regime by considering the description of the system in terms of an approximate Hamiltonian, as is shown in the following.


\section{Rokhsar Kivelson approximation}
\begin{figure}[t]
\centering
	\includegraphics[trim = 0mm 0mm 0mm 0mm, clip, width=\textwidth]{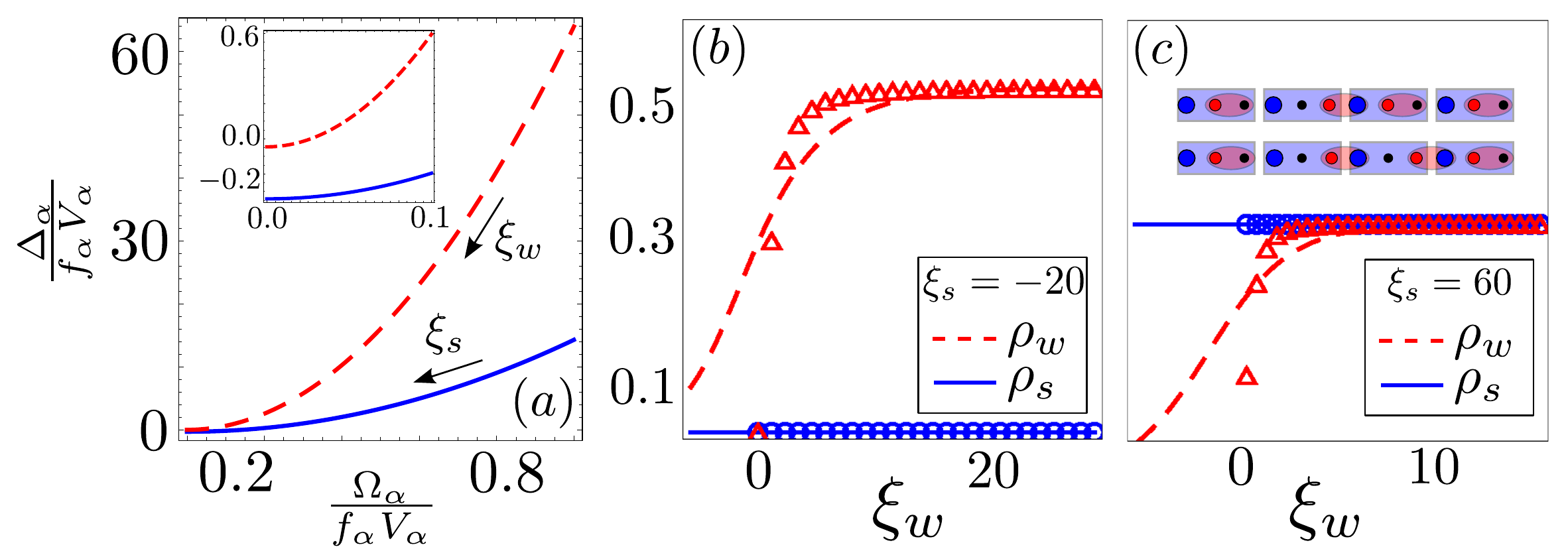}
\caption{(Color online)
(a) RK manifold of the $s$-atoms (solid blue line) and the $w$-atoms (red dashed line) \correction{for $V_s=100V_w=100$}. The RK manifold is parameterized by the parameters $\xi_\alpha$ which increase when going from right to left along the manifolds. \correction{The inset show a magnification of the main figure in the region of small $\Omega_\alpha$.} (b) Filling fractions $\rho_{\alpha}=(1/L)\sum_{k=1}^L \bra{G}n^{(\alpha)}_k\ket{G}$ as function of $\xi_{w}$ in the absence of $s$-atoms. Here $\rho_w$ saturates to its unfrustrated value $1/2$. \correction{The blue circles and red triangles represent the filling fractions obtained by diagonalizing exactly Hamiltonian (\ref{eq:hamiltonian}) for L=6 with periodic boundary conditions. We chose $V_s=100V_w=100$, and used the definitions of $\xi_{s,w}$ to fix $\Omega_{s,w}$, and Eq. (\ref{eq:RKmanifold}) to fix $\Delta_{s,w}$}. (c) In the presence of $s$-atoms the $w$-atoms can only assume a maximum filling fraction $\rho_s=1/3$ and form a frustrated state with \emph{exponentially many} configurations. The inset shows two example configurations with filling $\rho_w=1/3$. Here red ellipsoids (blue rectangles) sketch the dimers (trimers) of the RK construction. \correction{The blue circles and red triangles are the exact results drawn for comparison, and are obtained with the same method explained for panel (b)}.}
\label{fig:3}
\end{figure}

In the following we conduct an approximate analytical study of the ground state by means of a Rokhsar-Kivelson (RK) Hamiltonian $H_\mathrm{RK}$. This  generalizes the approach used in Refs. \cite{Lesanovsky11,Lesanovsky12,Levi} for characterizing the statics of a Rydberg gas.
The central idea is to approximate the Hamiltonian (\ref{eq:hamiltonian}) as a sum of local positive semi-definite projective Hamiltonians $H_{\mathrm{RK}}=\sum_{k=1}^{L}h_k$.
In this case a state $\ket{G}$ can be found which is annihilated by each of the local Hamiltonians, i.e. $h_k\ket{G}=0$, and thus represents the ground state of $H_\mathrm{RK}$. Under the RK approximation the ground state $\ket{G}$ is given by a superposition of classical configurations of a gas of hard-core polymers. As is shown below, this procedure neglects the long-range tails in the interactions. As a consequence the densities that the different species assume in the classical limit need to be adjusted by hand.

We consider a situation where in the classical limit the densities of the weak and strong species are given by $\rho_w=1/2$ and $\rho_s=1/3$. In this case the we can think of atoms in terms of polymers extending over sites at maximal distance $R_w=1$ (dimers) and $R_s=2$ (trimers), respectively (see Ref. \cite{Levi} for detail). The polymer analogy is equivalent to the hard-core conditions $n^{(w)}_k n^{(w)}_{k \pm 1}=0$ and $n^{(s)}_k n^{(s)}_{k \pm 1}=n^{(s)}_k n^{(s)}_{k \pm 2}=0$. Neglecting interactions on distances larger than $R_\alpha+1$ and making the hard-core conditions explicit also in the laser-excitation part of the Hamiltonian \cite{Lesanovsky11,Levi}, we can rewrite Hamiltonian (\ref{eq:hamiltonian}) as
\begin{eqnarray}
 \label{eq:apprHam}
 H = \sum_{k=1,\alpha=s,w}^L&& \left\{-\Omega_\alpha \xi_\alpha +\Omega_\alpha h^{(\alpha)}_k+\left[\Delta_\alpha-\Omega_\alpha\xi_{\alpha}^{-1}-\Omega_\alpha(2R_\alpha+1)\xi_{\alpha}\right]n_k^{(\alpha)}\right. \nonumber \\
 &&\left.+\left[\frac{V_\alpha}{(R_\alpha+1)^6}-\Omega_\alpha \xi_\alpha R_\alpha\right] n_k^{(\alpha)}n_{k+R+1}^{(\alpha)} \right\}.
\end{eqnarray}
where the local Hamiltonians $h^{(\alpha)}_k$ are given by
\begin{equation}
h^{(\alpha)}_k=\Pi^{(\alpha)}_k\left( \sigma^{x (\alpha)}_k+\xi_\alpha n^{(g)}_k+\xi^{-1}_\alpha n^{(\alpha)}_k \right).
\end{equation}
The $h^{(\alpha)}_k$ are the above-mentioned positive semi-definite operators which are proportional to projection operators and whose sum yields the RK Hamiltonian. The projection operators $\Pi^{(\alpha)}_k$ implement the hard-core condition by ensuring that dimers (trimers) do not occupy neighboring (next-neighboring) sites: $\Pi^{(w)}_k=p^{(w)}_{k-1}p^{(w)}_{k+1}$ and $\Pi^{(s)}_k=p^{(s)}_{k-2}p^{(s)}_{k-1}p^{(s)}_{k+1}p^{(s)}_{k+2}$ with $p^{(\alpha)}_k=1-n^{(\alpha)}_k$.

Hamiltonian (\ref{eq:apprHam}) is not yet of RK form since there are a number of additional terms. For specific parameter choices, however, these terms vanish up to trivial constants. In order to make this more transparent we have introduced (following Ref. \cite{Lesanovsky11}) the parameters $\xi_\alpha$. Those can be chosen freely since the Hamiltonian does not depend on them, as can be seen when multiplying out all the individual terms.

The RK form is assumed when choosing $\xi_\alpha=f_\alpha\,V_{\alpha}/\Omega_\alpha$ with $f_s=1/(2\times3^6)$ and $f_w=1/2^6$ and requiring the parameters $\Omega_\alpha, \Delta_\alpha, V_{\alpha}$ to accomplish
\begin{eqnarray}
\label{eq:RKmanifold}
      \left(\frac{\Omega_\alpha}{f_\alpha V_\alpha}\right)^2=\frac{\Delta_\alpha}{f_\alpha V_\alpha}+g_\alpha,
\end{eqnarray}
where $g_s=5$ and $g_w=3$. The first condition eliminates the two terms in the second line of Eq. (\ref{eq:apprHam}) while the second one leads to a cancellation of the terms proportional to $\Delta_\alpha$ and $\Omega_\alpha$. Both conditions define the so-called RK-manifold [see Fig. \ref{fig:3}(a)] on which Hamiltonian (\ref{eq:apprHam}) assumes the RK form
\begin{equation}
 \label{eq:Happrox}
 H_\mathrm{RK}=-(\Omega_s\xi_s+\Omega_w\xi_w)L+\sum_{k=1,\alpha=s,w}^L \Omega_\alpha h^{(\alpha)}_k.
\end{equation}

The ground state $\ket{G}$ of $H_\mathrm{RK}$ is written as
\begin{equation}
 \label{eq:GS}
  \ket{G}=\prod_{k=1}^N\left(1-\xi_d \Pi^{(w)}_k c^{(w)\dagger}_k-\xi_t \Pi^{(s)}_k c^{(s)\dagger}_k\right)\ket{0},
\end{equation}
where $\ket{0}=\bigotimes_{i=1}^N\ket{g}_i$ and $c^{(\alpha)\dagger}=\ket{\alpha}\bra{g}$.
The parameters $\xi_\alpha$ can be then thought of as the fugacities of the dimers and trimers gases, such that large (small) values result in a ground state with high (low) density of the respective species.
The ground state (\ref{eq:GS}) can be represented explicitly as matrix product state (see e.g. \cite{Levi}). We can now use the ground state $H_\mathrm{RK}$ to get an idea of the effect of quantum fluctuations on crystalline states. We find that frustrated states emerge also away from the classical limit, i.e. for non-zero $\Omega_\alpha$.

By construction $H_\mathrm{RK}$ is frustration free, i.e. the frustration energy (\ref{eq:fe}) is always zero. In fact this is an artifact stemming from the omission of long-range tails. Nevertheless, the interspecies frustration still manifests in the behaviour of the filling fractions, Fig. \ref{fig:3}. In Fig. \ref{fig:3}(b), where we set $\xi_s<0$ in order to have a low density of dimers, we find that indeed only dimers are occupying the lattice and that they fill it completely in the limit $\xi_w \rightarrow \infty$. This corresponds to a unique (translationally invariant) crystalline ground state with $\rho_w = 1/2$. On the other hand in Fig. \ref{fig:3}(c) we fix the parameter $\xi_s$ such that the strong species (trimers) fills up the lattice at density  $\rho_s=1/3$. We now increase $\xi_w$ which increases the density of the weak species. However, instead of a saturation at density $1/2$ --- as in the absence of the strong species --- we find that $\rho_w$ tends to a value of $1/3$. This is a signature of inter-species frustration. The weak species does not crystallize but is in a superposition of exponentially many states with density $1/3$, two of which are shown in the inset of Fig. \ref{fig:3}(c).


\section{Summary and outlook}
We have investigated the statics of a two component Rydberg gas at zero temperature, with particular focus on the emergence of frustration effects. In the classical regime we identified regions in parameter space where the two Rydberg species form compatible and incompatible arrangements. In the quantum case we introduced the frustration energy to quantify the degree of frustration. We found that this quantity shows a staircase pattern as the system visits different phases characterized by different filling fractions. Finally, we performed an approximate analytical study of the quantum regime by means of an RK Hamiltonian, which further corroborated the existence of frustrated states.

Our analysis provides insights into the complexity of ground state structures in multi-component Rydberg gases. In the future it would be interesting to extend the work to situations without the strong-weak hierarchy that was assumed here. Moreover, an extension of the analysis to Rydberg gases with more than two relevant excited states and/or to high-dimensional lattices would be desirable due to the number of Rydberg lattice experiments that recently has become available.

\section{Acknowledgements}
We are very indebted to Weibin Li for providing us with data used to plot Fig. \ref{fig:X}. E.L. would like to thank M. Marcuzzi for insightful discussions. J. M. would like to thank H. Weimer for useful discussions. The research leading to these results has received funding from the European Research Council under the European Union's Seventh Framework Programme (FP/2007-2013) / ERC Grant Agreement No. 335266 (ESCQUMA), the EU-FET grants HAIRS 612862 and from the University of Nottingham.  Further
funding was received through the H2020-FETPROACT-2014 Grant No.  640378 (RYSQ) and the EPSRC Grant no.\ EP/M014266/1. \\
\\
\\

\bibliographystyle{iopart-num}

\providecommand{\newblock}{}

\end{document}